# Extraction of the Solvation Structure on a Solid Plate from a Force Curve Measured by Surface Force Apparatus in a Hard-Sphere Fluid

## Ken-ich Amano[1] and Ohgi Takahashi[2]

[1] *Department of Energy and Hydrocarbon Chemistry, Graduate School of Engineering, Kyoto University, Kyoto, 615-8510, Japan.*

[2] *Faculty of Pharmaceutical Sciences, Tohoku Pharmaceutical University, 4-4-1 Komatsushima, Aoba-ku, Sendai 981-8558, Japan.*

Author to whom correspondence should be addressed: Ken-ichi Amano.

Electric mail: amano.kenichi.8s@kyoto-u.ac.jp

**ABSTRACT**

We propose a method of transformation from a force curve obtained with a surface force apparatus (SFA) to a density distribution of a liquid near a surface. The method is based on the statistical mechanics of liquids. As a first step, we show the method for a rigid system in which two cylinders are immersed in a hard-sphere fluid as the force probes. We found that the method works well, especially in the lower density solvent. The accuracy of the transformed result increases as the distance between the circular surface of the cylinder and the solvent sphere increases.

**MAIN TEXT**

The surface force apparatus (SFA) has been utilized to measure the force between two surfaces in solvents [1,2]. The force curve shows oscillating behavior, and it is therefore called the oscillation force. The origin of the oscillation is due to the confinement of the solvent particles between the two surfaces, as elucidated by simulations [3,4]. Moreover, the theoretical relationship between "the force curve" and "the pair correlation function between the two surfaces" has already been studied [5]. Specifically, the force curves obtained by the SFA measurements have been well understood. One of the next topics in SFA research is a method of transformation from the force curve to the density distribution of a liquid near a surface—the "inverse calculation". Thus, in the present paper, we propose a method based on the statistical mechanics of liquids. We first apply the method to a rigid system where the two cylinders are immersed in a hard-sphere fluid as the force probes. Since the transformation method has not been theoretically derived even in the simple system, the method could become the groundwork for the inverse calculation of SFA. After the derivation of the method, we describe verification tests of the method using a three-dimensional (3D) integral equation theory.

Recently, Amano *et al*. [6] proposed a method for transforming the force distribution [7,8] measured by atomic force microscopy (AFM) in a liquid to the intrinsic solvation structure on a solid plate. This method is constructed based on the statistical mechanics of liquids. Since the method is related to the SFA system, we take advantage of the method for the inverse calculation of the SFA. Here, to start the derivation we introduce six conditions:

(I) Cylindrical solids 1 and 2 of the rigid bodies are immersed in a hard-sphere fluid (see FIG. 1). The cylindrical solids have the same shapes. (The cylindrical solids can be alternated with cuboidal solids in this theory. However, we use only the cylindrical solids here to make this explanation simple.)

(II) The circular surfaces of the cylindrical solids are facing each other, and the facing surfaces are vertical to the $z$-axis.

(III) The origin of the system is set at the center of the facing surface of cylindrical solid 1. Cylindrical solid 1 is fixed, whereas cylindrical solid 2 changes its position only along the $z$-axis.

(IV) The facing surface areas are sufficiently large to ignore influences of the corners of the cylindrical solids, i.e. the forces reproduced by solvated particles on the corners are approximated to *zero*.

(V) The lateral surfaces of the cylindrical solids (they are not the facing surfaces) are horizontal to the $z$-axis, so that no force along the $z$-axis is generated from the solvated particles on the lateral surfaces.

(VI) The thicknesses (lengths) of the cylindrical solids are sufficiently large, so that the solvation structures on the reverse (circular) surfaces are never destroyed during any separation between the facing surfaces.

In this case, the force along the $z$-axis acting on cylindrical solid 2 ($f$) is expressed as follows:

$$f(s) = A \int_{-\infty}^{\infty} \rho(z;s) \frac{\partial u_2(z;s)}{\partial z} dz, \tag{1}$$

where $A$ represents the facing surface area of cylindrical solid 2. $\rho(z;s)$ is the number density of the solvent at $z$, and $s$ is the separation between the facing surfaces. $u_2$ is the two-body potential between cylindrical solid 2 and the solvent particle. Eq. (1), derived by considering an infinitesimal movement of the cylindrical solid 2 in the system, is an exact one in the classical statistical mechanics of liquids. This equation is strictly consistent with the contact theorem [9-12]. (The contact theorem explains the pressure on a wall, the derivation of which is performed by an infinitesimal change of the volume of a system or a solute.) To connect the force ($f$) and the solvation structure on cylindrical solids 1 and 2 ($g_1$ and $g_2$), we take advantage of the Kirkwood superposition approximation [13] and express $\rho$ as $\rho_0 \cdot g_1 \cdot g_2$ [14,15], where $\rho_0$ is the bulk number density of the solvent (which is constant). The solvation structure $g$ is called the pair correlation function between the cylindrical solid and the solvent particle or the normalized liquid density. Using this approximation, Eq. (1) is rewritten as

$$f(s) = A\rho_0 \int_{-\infty}^{\infty} g_1(z) g_2(z-s) \frac{\partial u_2(z-s)}{\partial z} dz. \tag{2}$$

Here, we set the origin of the function $g_1$ at the center of the facing surface of cylindrical solid 1 (i.e., the origin of the whole system), whereas the origins of the functions $g_2$ and $u_2$ are set at the center of the facing surface of cylindrical solid 2. Considering the conditions (V) and (VI), Eq. (2) can be rewritten as

$$f(s) = A\rho_0 \int_{0}^{s} g_1(z) g_2(z-s) \frac{\partial u_2(z-s)}{\partial z} dz + A\rho_0 \int_{s+w}^{\infty} g_2(z-s) \frac{\partial u_2(z-s)}{\partial z} dz, \tag{3}$$

where $w$ represents the thickness (length) of cylindrical solid 2 (i.e., the distance between the facing surface and the reverse surface of cylindrical solid 2). In this derivation, the two-body potential between cylindrical solid 2 and the solvent particle $u_2$ is expressed as

$u_2 = 0$      for contact points and non-overlapped points,          (4a)

$u_2 = \infty$      for overlapped points.          (4b)

Then, the value of $\exp[-u_2/(k_B T)]$ is expressed as

$\exp[-u_2/(k_B T)] = 1$      for contact points and non-overlapped points,      (5a)

$\exp[-u_2/(k_B T)] = 0$      for overlapped points,      (5b)

where $k_B$ and $T$ are the Boltzmann constant and absolute temperature, respectively. Thus, the partial differentiation of $u_2$ with respect to $z$ can be expressed as

$$\frac{\partial u_2(z-s)}{\partial z} = k_B T \exp[u_2(z-s)/(k_B T)]\{\delta[z-(s-d_S/2)] - \delta[z-(s+w+d_S/2)]\},\ (6)$$

where $\delta$ and $d_S$ are the delta function and the diameter of the solvent particle, respectively. By substituting Eq. (6) into Eq.(3), the force acting on cylindrical solid 2 is written as

$$f(s) = A k_B T \rho_0 g_1(s-d_S/2) g_2(-d_S/2) - A k_B T \rho_0 g_2(w+d_S/2). \quad (7)$$

Since the values of $g_2(-d_S/2)$ and $g_2(w+d_S/2)$ are for the contact points, they are both written as $g_C$, where the subscript C identifies the contact point. Hence, the force is rewritten as

$$f(s) = A k_B T \rho_0 g_c [g_1(s-d_S/2) - 1]. \quad (8)$$

When $s = d_S$, $g_1(s-d_S/2)$ is $g_1(d_S/2)$, the value of which is equal to $g_C$. Therefore, the value of $g_C$ can be calculated as

$$g_c = \frac{1}{2} + \frac{\sqrt{A^2 k_B{}^2 T^2 \rho_0{}^2 + 4 A k_B T \rho_0 f(d_S)}}{2 A k_B T \rho_0}. \quad (9)$$

In the derivation of Eq. (9), we used the fact that $g_C$ is equal to 1 when $f(d_S)$ is 0. This equation implies that if $f(d_S)$ is measured (under the six conditions), the contact density of the hard-sphere fluid ($\rho_0 \cdot g_C$) can be obtained through Eq. (9). As mentioned above, if $f(d_S)$ is equal to 0, $g_C$ becomes 1. However, $g_C$ is generally larger than 1 in the rigid system. Hence, we predict that $f(d_S)$ is larger than 0 under the six conditions, i.e., a repulsive force is generated when the separation between the *flat* surfaces is equal to the diameter of the solvent particle. Since $g_C$ is obtained through Eq. (9), $g_1$ can be calculated as follows:

$$g_1(s - d_S/2) = \frac{f(s)}{Ak_BT\rho_0 g_c} + 1 \qquad \text{where } s \geq 0. \tag{10}$$

Here, let us consider the value of $g_1$ within the range of $0 \leq s < d_S$ under the six conditions. In this case, the force acting on cylindrical solid 2 comes only from its reverse surface because there are no solvent particles between the facing surfaces. The force from the reverse surface $f_B$ (constant) can be obtained from the second term of the right-hand side of Eq. (7):

$$f_B(s) = -Ak_BT\rho_0 g_C. \tag{11}$$

When $0 \leq s < d_S$, $f$ is equal to $f_B$, and therefore, the value of $g_1(s - d_S/2)$ is calculated as 0 by substituting Eq. (11) into (10). The behavior of $g_1$ is physically valid.

In what follows, we examine the transformation method by using a 3D integral equation theory [16,17]. We first use the 3D Ornstein-Zernike equation coupled with a hypernetted-chain closure (3D-OZ-HNC) to calculate the solvation structure around the single cylindrical solid $g_0$. (The aim of this examination is a comparison of $g_0$ and $g_1$.) We then calculate the solvation structure around the two cylindrical solids with an arbitrary separation (see FIG. 1) using 3D-OZ-HNC. Next, we use the 3D Singer-Chandler formula to calculate the solvation free energy (SFE) of the two cylindrical solids with an arbitrary separation. We then calculate the force acting on cylindrical solid 2 ($f$) by partial differentiation of the SFE and transform $f$ into $g_1$ through Eq. (10). Finally, we compare $g_0$ and $g_1$ for the four solvent densities. As shown in FIG. 2, the method works well, especially in the lower density solvents. The accuracy of the transformed result ($g_1$) becomes higher as the distance between the circular surface of the cylindrical solid and the solvent sphere increases. Therefore, the method we proposed here is considered to be valid in theory.

The verification tests above showed that the method could be applicable to SFA experiments. However, there were *amplitude* differences between $g_0$ and $g_1$, especially near the contact point. In our opinion, this problem can be overcome by the introduction of a modified Kirkwood superposition approximation. For instance, the modified one would be written as $\rho \approx \rho_0(g_1 \cdot g_2)^\alpha$, where $\alpha$ is a modification parameter. The value of $\alpha$ is around 1, and it can be estimated by calculations and/or experiments.

Furthermore, there were *periodic* differences between $g_0$ and $g_1$, especially in the higher density solvents. The difference originates from the confinement of the solvent particles between two *large* surfaces. Since the surface of the SFA probe is very large compared with the AFM, the confined solvent particles tend to be significantly compressed. This compression is the reason for the *periodic* differences. The Kirkwood superposition approximation cannot treat the compression behavior, and thus, another type of approximation is needed to make the inverse calculation more accurate. (Fortunately, liquid AFM theory indicates [18,19] that if a probe tip is sufficiently small and the solvation affinity of the probe surface is similar to the solvent particle, the compression could be ignored in some degree in the AFM research.) In order to overcome such a deviation, we suggest an approximation below:

$$\rho(s - d_S/2) = \rho_0 g_1(s - d_S/2)g_2(-d_S/2)/\exp[-Q(s)/C], \qquad (12)$$

where $Q$ represents surface free energy per unit area of the flat surface and $C$ is a parameter. This approximation is similar to both Kirkwood superposition approximation [13-15] and Percus shielding approximation [20-22]. In the hard-sphere fluid with $\rho_0 d_S^3 = 0.7$, we determined that C is about 0.01 and found that Eq. (12) works well for inverse calculation of $g_1$ in the hard-sphere fluids with $\rho_0 d_S^3 = 0.4$, 0.5, 0.6, and even 0.8. (The results are not shown here, and we notify the forms of Eqs. (9) and (10) are slightly changed when Eq. (12) is used.)

In summary, we have proposed a method for transforming a force curve measured by SFA under six conditions into a solvation structure on a flat solid. The rigid cylindrical solids were immersed in the hard-sphere fluid, and the derivation was performed based on the statistical mechanics of liquids. This transforming method is applicable when the experimental condition (almost) meets the six conditions we set. Concerns about the method are the introduction of the Kirkwood superposition approximation and the existence of a phase transition upon the confinement of a liquid by two large surfaces. If the approximation fails to work in an actual system or if the phase transition occurs, the transformation method must not be used. In the near future, we plan to demonstrate the transform methods coupled with the modified superposition approximations. Moreover, we are going to demonstrate the inverse calculation in non-rigid system (e.g., the system with Lennard-Jones potentials).

## ACKNOWLEDGEMENTS

We appreciate Masahiro Kinoshita (Kyoto University), Ryo Akiyama (Kyushu University), Hiroshi Onishi (Kobe University), and Kota Hashimoto (Kyoto University) for useful discussions. This work was supported by the "Foundation of Advanced Technology Institute" and "Joint Usage/Research Program on

Zero-Emission Energy Research, Institute of Advanced Energy, Kyoto University (ZE25B-30)."

**FIGURES**

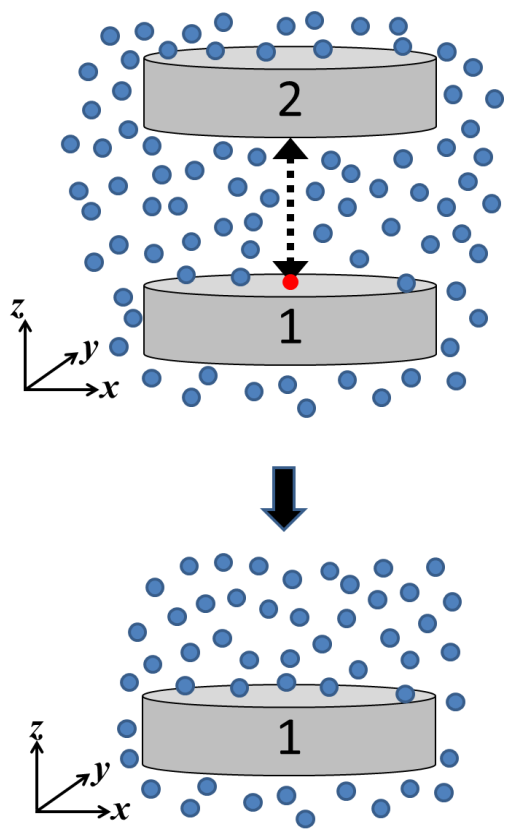

FIG. 1. Cylindrical solids 1 and 2 are immersed in the hard-sphere fluid. The red point is the origin of the entire system. The center of the circular surface of cylindrical solid 1 is fixed at the origin. The dotted arrow represents the separation between the two circular surfaces. At first, the force acting on cylindrical solid 2 is determined. Then, the solvation structure on cylindrical solid 1 is calculated by the transforming method.

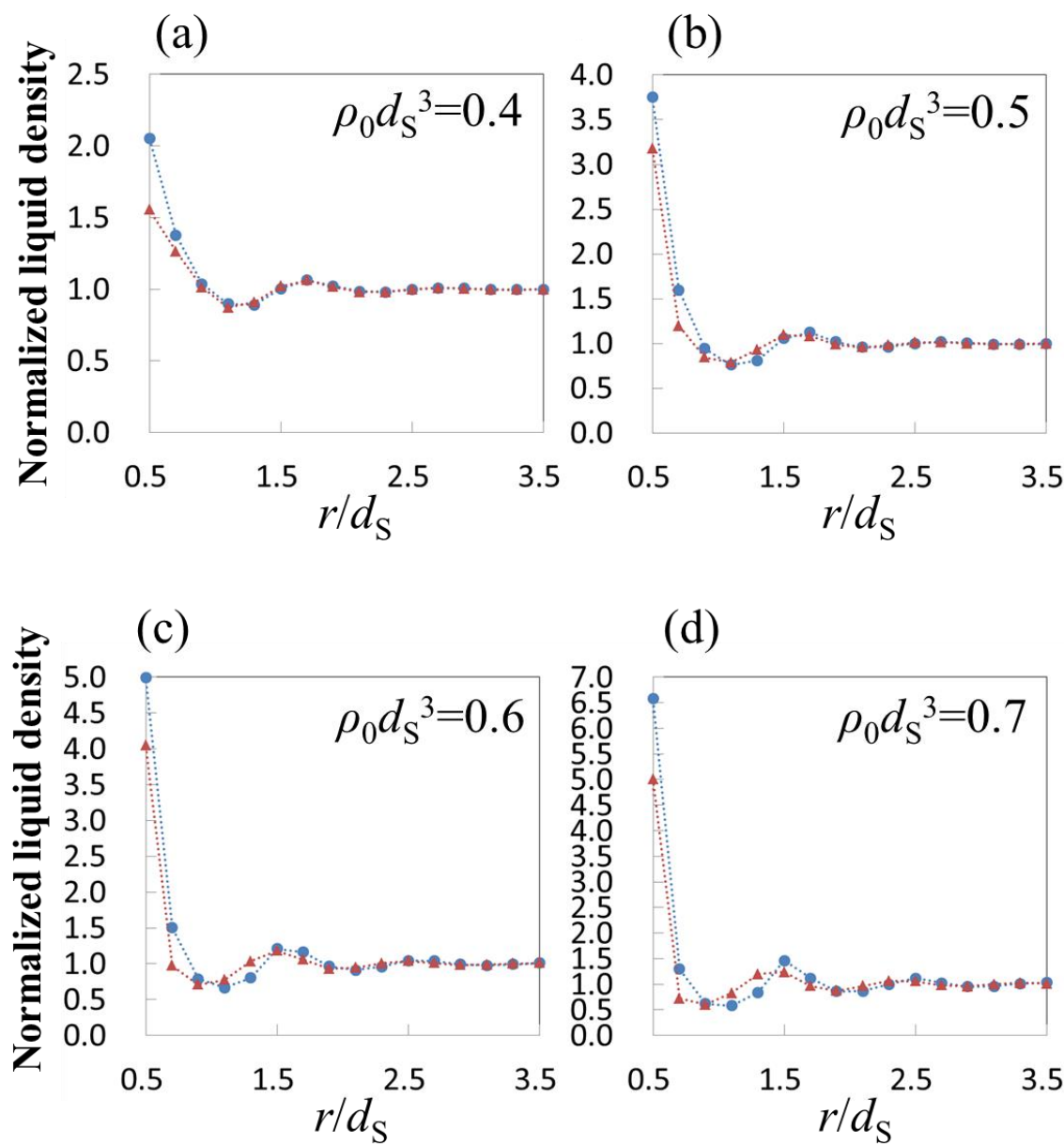

FIG. 2. Comparisons between the solvation structures calculated by the normal process ($g_0$) and the transformation method ($g_1$). $g_0$ and $g_1$ are represented by circular and triangular points, respectively. Dimensionless values of the solvent density ($\rho_0 d_S^3$) are 0.4, 0.5, 0.6, and 0.7 in (a), (b), (c), and (d), respectively. $r$ represents the distance between the circular surface and the center of the solvent sphere. When $r/d_S = 0.5$, the solvent sphere contacts the circular surface. (For calculation of $g_1$, Eqs. (9) and (10) are used.)

1<sup>st</sup> Submission: 3 July (2013) EST.

2<sup>nd</sup> Submission: 10 July (2013) EST.
Ohgi Takahashi is added as a coauthor; Eqs. (5a) and (5b) are added; English expressions are improved.

3<sup>rd</sup> Submission: 29 July (2013) EST.
Ryo Akiyama and Hiroshi Onishi are added in the **ACKNOWLEDGEMENTS**.

4<sup>th</sup> Submission: 12 Dec (2013) EST.
Calculation results and its related materials and sentences are added.

5<sup>th</sup> Submission: 16 November (2015) EST.
Affiliation and e-mail address are renewed, an approximation (Eq. (12)) that is better than the Kirkwood superposition approximation is suggested, references [20-22] are added, and Kota Hashimoto is added in the **ACKNOWLEDGEMENTS**.